\documentclass[usegraphicx,usenatbib]{mn2e}

\usepackage{url}
\usepackage[breaklinks=true]{hyperref}
\usepackage{twoopt}
\usepackage[english]{babel}          % English language/hyphenation

\usepackage{natbib}
\bibpunct{(}{)}{;}{a}{}{,} %% natbib format for A&A and ApJ

\usepackage[utf8]{inputenc}
\usepackage[normalem]{ulem}
\usepackage{siunitx}
\usepackage{amsmath, amssymb}
\usepackage[T1]{fontenc} % MNRAS suggestion for combining PDF on submission
\usepackage{aecompl} % MNRAS suggestion for combining PDF on submission
\usepackage[farskip=0pt]{subfig}

\usepackage{multirow,bigstrut,ctable}
\usepackage[normalem]{ulem}
\usepackage{rotating}
\usepackage[varg]{txfonts} % A&A recommended fonts
\usepackage{threeparttable}

\def \deg         {\text{$^{\circ}$}}
\def \arcmin      {\text{$^\prime$}}
\def \arcsec      {\text{$^{\prime\prime}$}}
\def \hour        {$^{\mathrm{h}}$}
\def \min         {$^{\mathrm{m}}$}
\def \sec         {$^{\mathrm{s}}$}

\def \mujybeam    {$\mu$Jy\,beam$^{-1}$}

\def \radmm       {rad\,m$^{-2}$}

\newcommand{\hms}[3]{{#1}\hour{#2}\min{#3}\sec}
\newcommand{\dms}[3]{{#1}\deg{#2}\arcmin{#3}\arcsec}
\newcommand{\beam}[2]{{#1}\arcsec$\times${#2}\arcsec}

\def \targetA       {NVSS J232149+484951}
\def \targetB       {NVSS J232147+482956}

\title[Multifrequency study of a new HyMoRS]
      {Multifrequency study of a new Hybrid Morphology Radio Source}

\author[F.~de~Gasperin]{F. de Gasperin$^{1,2}$
\\
$^{1}$ Leiden Observatory, Leiden University, P.O.Box 9513, NL-2300 RA, Leiden, The Netherlands\\
$^{2}$ Hamburg Observatory, University of Hamburg, Gojenbergsweg 112, D-21029, Hamburg, Germany}

\begin{document}

\date{}
\pagerange{\pageref{firstpage}--\pageref{lastpage}} \pubyear{2013}
\maketitle

\label{firstpage}

\begin{abstract}
Hybrid Morphology Radio Sources (HyMoRS) are a class of radio galaxies having the lobe morphology of a Fanaroff-Riley (FR) type I on one side of the active nucleus and of a FR type II on the other. The origin of the different morphologies between FR\,I and FR\,II sources has been widely discussed in the past 40 years, and HyMoRS may be the best way to understand whether this dichotomy is related to the intrinsic nature of the source and/or to its environment. However, these sources are extremely rare ($\lesssim 1\%$ of radio galaxies) and only for a few of them a detailed radio study, that goes beyond the morphological classification, has been conducted. In this paper we report the discovery of one new HyMoRS; we present X-ray and multi-frequency radio observations. We discuss the source morphological, spectral and polarisation properties and confirm that HyMoRS are intrinsically bimodal with respect to these observational characteristics. We notice that HyMoRS classification based just on morphological properties of the source is hazardous.
\end{abstract}


\section{Introduction}

\cite{Fanaroff1974} divided radio galaxies into two categories based on their morphology. Type\,I have slower, more turbulent, and less collimated jets. Conversely, type\,II have powerful collimated jets (often only one is visible) which terminate into bright, compact ``hot spots''. In these regions much of the bulk energy of the jet is converted into accelerating relativistic particles and amplifying magnetic fields through shocks produced by the beam against the intergalactic medium \citep{Blandford1974, Carilli1991}. While the FR\,I class is made up of sources with different morphologies, the majority of them show jets that appear laminar in the innermost region inflating turbulent lobes after passing through a flare point. This characteristic can be well explained by the interaction with the ambient medium \citep{Laing2014}. The magnetic field orientation for FR\,Is is often perpendicular to the jet, apart from the innermost region. Conversely, FR\,II jets have a magnetic field parallel to the jet orientation \citep{Bridle1984}.

Another distinction between FR\,I and FR\,II was based on jet/lobe luminosity, with FR\,I having lower luminosity ($<10^{25}$ W Hz$^{-1}$ at 1.4 GHz) than FR\,II. The value of the radio luminosity transition depends on the properties of the host galaxies as it tends to increase with the host optical luminosity \citep{Ledlow1996a}, but this result has not been confirmed by subsequent analysis \citep{Best2009, Wing2010}. This has been interpreted as a link between the morphological dichotomy of radio galaxies and the properties of either the nucleus \citep[e.g.][]{Ghisellini2001} or the environment \citep[e.g.][]{Bicknell1995}. FR\,Is are found in richer clusters \citep{Prestage1988} and this relation evolves in time, while at high-$z$ FR\,IIs are also found in cluster centres \citep{Hill1991}.

A peculiar sub-class of radio galaxies have a hybrid FR\,I/FR\,II morphology. Hybrid morphology radio sources (HyMoRS) were first defined by \cite{Gopal-Krishna2000} as a ``class of double radio sources in which the two lobes exhibit clearly different Fanaroff-Riley morphologies''. \cite{Gawronski2006} inspected 1700 sources from the FIRST survey finding three certain and two possible HyMoRS showing that these type of sources are rare ($\lesssim 1\%$ of radio galaxies belong to this category). Currently, around 10 HyMoRS are considered genuine but only a small fraction of them have been studied in detail. \cite{Pirya2011} made a detailed multi-wavelength radio study of two possible HyMoRS (J1211+743 and J1918+742). While, in other wavelengths, \cite{Miller2009} made \textit{Chandra} observations of the two hybrids 3C\,433 and 4C\,65.15. They found that both sources have unabsorbed X-ray luminosity, radio luminosity, and optical spectra of a typical FR\,II, while the FR\,I side is likely a consequence of jet-medium interaction. Finally, \cite{Ceglowski2013} examined the central 10 kpc of five known HyMoRS using VLBI observations. Their work underlines that genuine HyMoRS are not a consequence of the jet orientation and that on 1--10~kpc scale their targets have weak jets, compatible with FR\,II-like structures. The study of hybrid radio sources might be of fundamental importance to understand the origin of the FR dichotomy and assess whether the large scale morphology of radio galaxies is linked to the central engine property, the environment or to a combination of the two.

Even though recent efforts have been undertaken to find hybrid sources, the known population of HyMoRS is still small and, to our knowledge, no study has so far combined high-resolution, polarisation and spectral index radio maps to characterise a candidate HyMoRS. In this paper we report the discovery of a new HyMoRS: \targetA{} (hereafter RG1). In Sec.~\ref{sec:analysis} we present multi-frequency observations of the source and we analyse high-resolution and polarisation radio maps. In Sec.~\ref{sec:rg1} we present the newly discovered HyMoRS and we make spectral index characterisation in Sec.~\ref{sec:spidx}. In Sec.~\ref{sec:rg2} we discuss another radio galaxy (\targetB{}, hereafter RG2) as an example of the difficulties in the HyMoRS classification when using only morphological data. Discussion and conclusions are in Sec.~\ref{sec:conclusions}.

Throughout the paper we adopt a fiducial $\Lambda$CDM cosmology with $H_0 = 70\rm\ km\ s^{-1}\ Mpc^{-1}$, $\Omega_m = 0.3$ and $\Omega_\Lambda = 0.7$. At the redshift of the target ($z\approx0.16$) 1\arcsec = 2.8 kpc. All regressions are made with a bootstrap least squares algorithm that takes into account errors on the dependant variables. Unless otherwise specified errors are at $1\sigma$. The spectral index is defined as: $F_{\nu} \propto \nu^\alpha$, where $F_\nu$ is the flux density.

\section{Observations and data reduction}
\label{sec:analysis}

\subsection{Radio}
We obtained GMRT data at 323 and 607 MHz, WSRT data at 1380 MHz and VLA data at 1--2 GHz (see Table~\ref{tab:obs}). The observations are pointed towards the nearby galaxy cluster PSZ1 G108.18-11.53 ($z=0.335$), but both radio galaxies described in this paper fall well within the primary beam full width half maximum of these observations. The data reduction procedures for the GMRT and WSRT observations are described in \cite{deGasperin2015a} and take advantage from the SPAM (Source Peeling and Atmospheric Modeling) package for ionospheric calibration to achieve high-fidelity, thermal noise limited images \citep{Intema2009}.

\begin{table*}
\centering
\begin{threeparttable}
\begin{tabular}{lccccccc}
Telescope & Frequency & Obs  & Total observing time & Bandwidth & Max resolution & Rms noise   & Figure \\
          & (MHz)     & Date & (hours)              & (MHz)     & (arcsec)       & (\mujybeam) & \\
\hline
GMRT & 323 & 02 Jun 14 & 9 & 32 & \beam{10.5}{8.2} & 92 & \ref{fig:radio} Top-left\\
GMRT & 607 & 06 Jun 14 & 8 & 32 & \beam{5.4}{5.3} & 47 & \ref{fig:radio} Top-right\\
WSRT & 1381 & 02 Jan 14 & 8 & 160 & \beam{17.3}{13.3} & 32 & \ref{fig:radio} Bottom-left\\
VLA & 1519 & 02/05 Apr 15; 30 Jan 16 & 10\tnote{a} & 1000 & \beam{4.8}{4.3} & 10 & \ref{fig:radio} Bottom-right\\
\end{tabular}
\begin{tablenotes}
\item[a] 6 h in B-configuration and 4 h in C-configuration
\end{tablenotes}
\end{threeparttable}
\caption{Radio observations}\label{tab:obs}
\end{table*}

\begin{figure*}
\centering
\includegraphics[width=.49\textwidth]{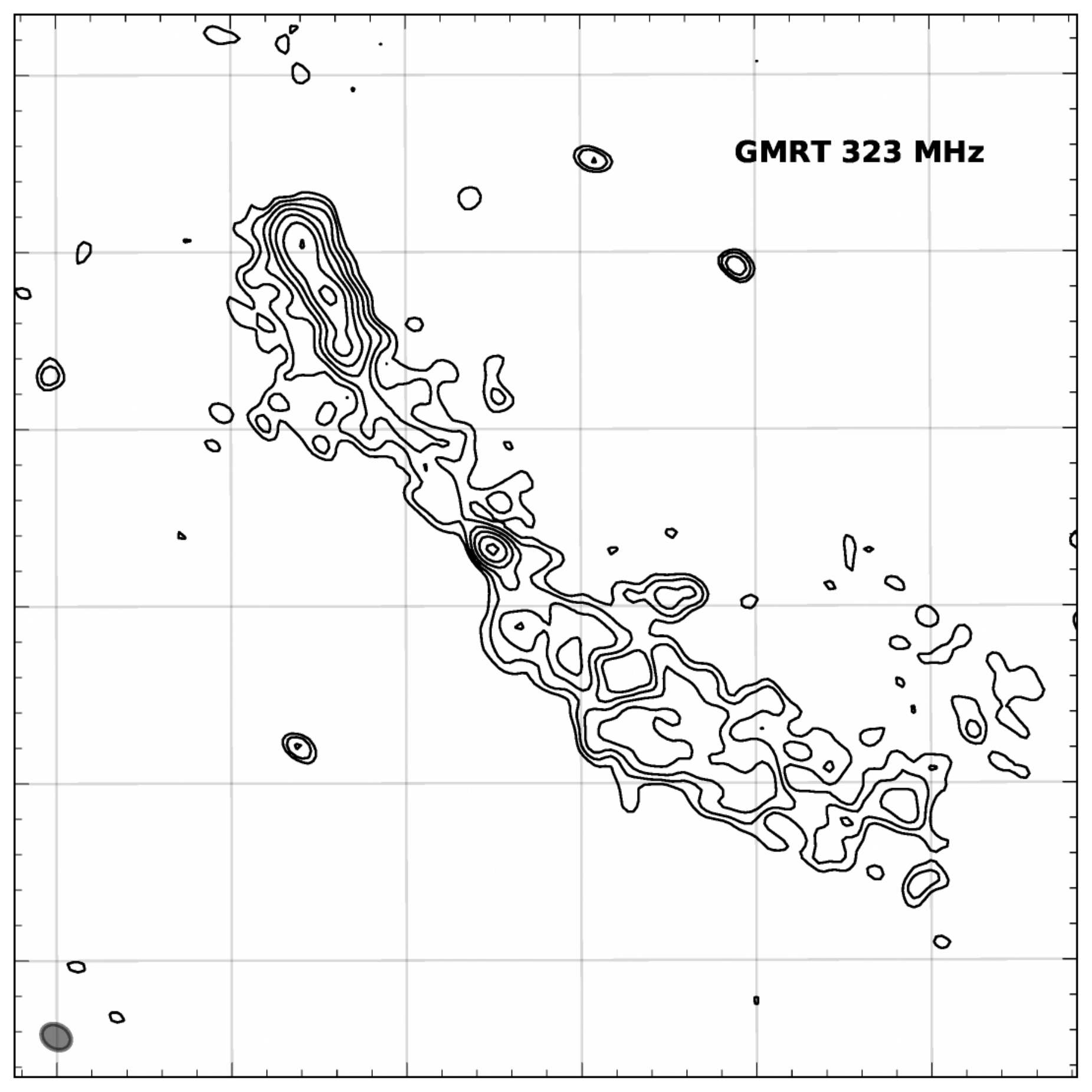}
\includegraphics[width=.49\textwidth]{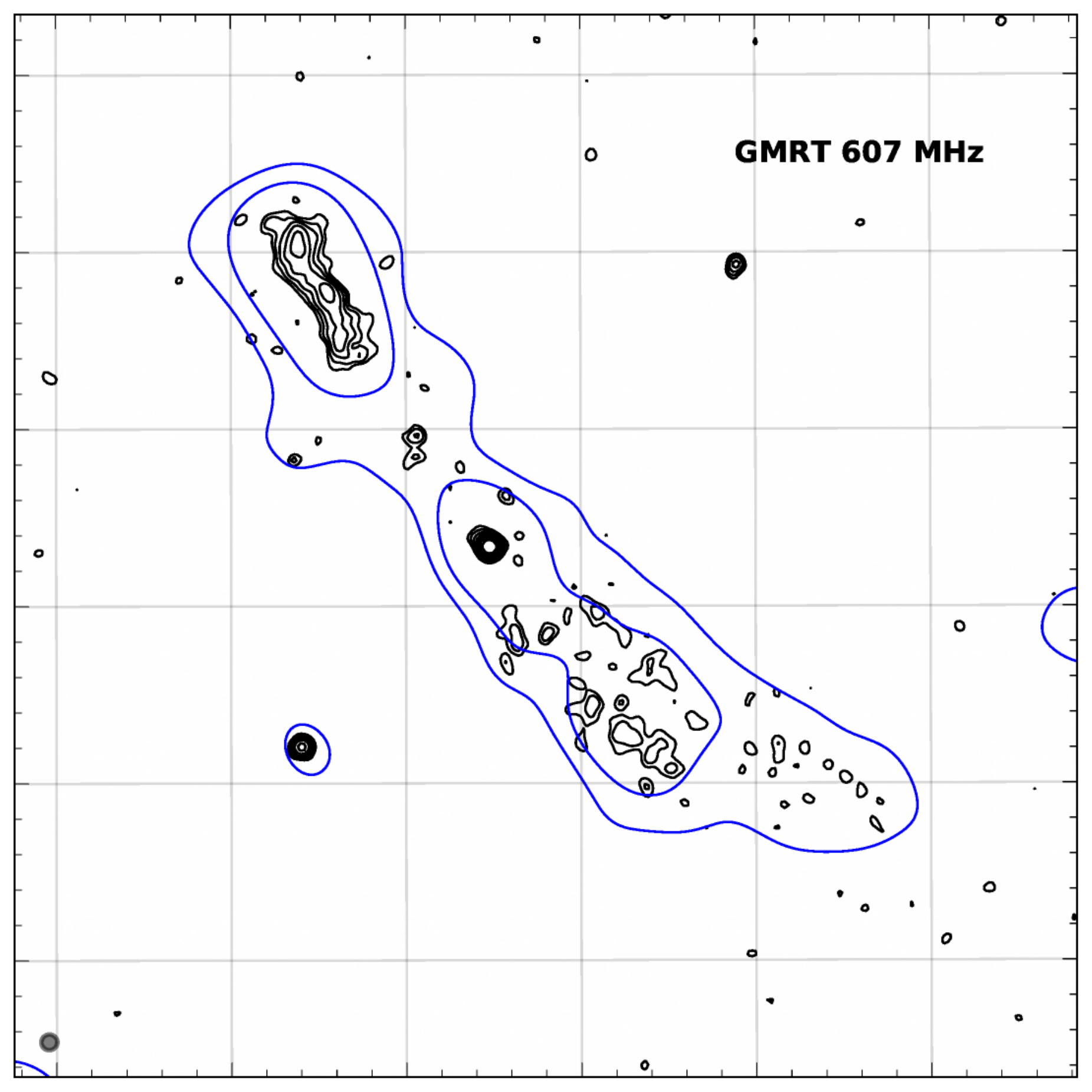}\\
\includegraphics[width=.49\textwidth]{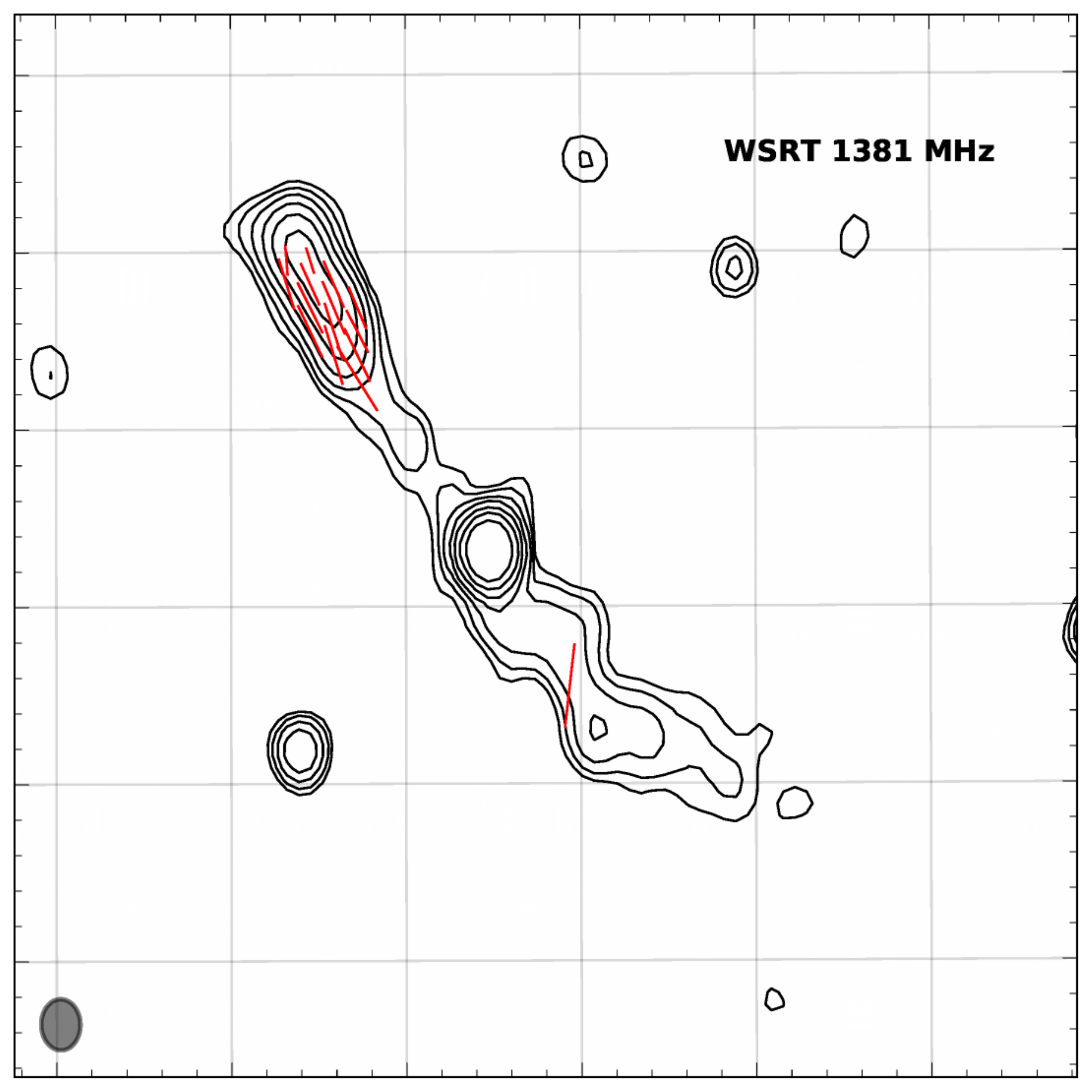}
\includegraphics[width=.49\textwidth]{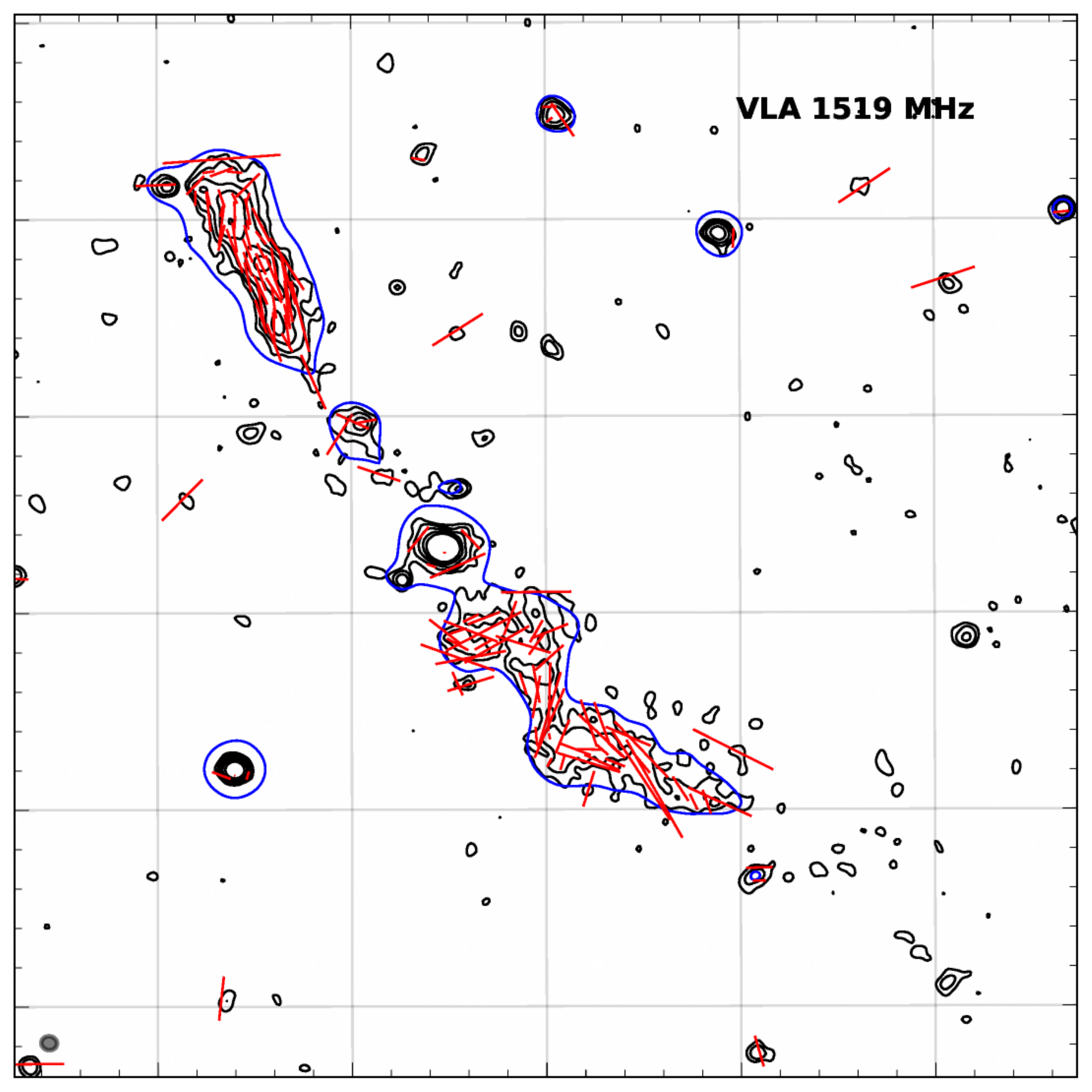}\\
\caption{Black contours are from GMRT/WSRT/VLA data (323, 607, 1381, and 1519 MHz, as labelled in the images) at $3...30\times\sigma$ with sigma and beam size as described in Table~\ref{tab:obs}. Blue contours are $(3,6)\times\sigma$ from a low-resolution images ($\sigma = 170$~\mujybeam{}, beam = \beam{18}{18} for GMRT 607 MHz and $\sigma = 15$~\mujybeam{}, beam = \beam{11}{10} for VLA 1519 MHz). Polarisation vectors show the B-mode (parallel to magnetic field orientation) with a length proportional to the fractional polarisation which, for the VLA image, is on average 18\% in the northern lobe and 35\% at the end of the southern lobe. Polarisation angle has been corrected for Galactic rotation which is this region of the sky is expected to be $-14$~\radmm \citet{Oppermann2012}.} \label{fig:radio}
\end{figure*}

VLA data has been taken in the L-band (1--2 GHz) in B and C configuration. The data were reduced using the CASA\footnote{https://casa.nrao.edu/} package. The visibilities were Hanning-smoothed, bandpass-calibrated and flagged using the automatic tool AOflagger \citep{Offringa2012}. We used 3C147 as flux calibrator and 3C138 to calibrate the polarisation angle. The flux scale has been set to \cite{Perley2013} which above 1 GHz is in-line with \cite{Scaife2012} used for the low frequency observations. Bandpass, scalar delays, cross-hand delays, and polarisation angle corrections were transferred to the target and the phase calibrator. Then, phase and rescaled amplitude from the phase calibrator were transferred to the target field. Finally, a single cycle of phase-only self-calibration was applied on the target field.

We note that all flux density errors for extended emission are computed as $S_{\rm err} = \sigma \cdot \sqrt{N_{\rm beam}}$, where $\sigma$ is the local image rms and $N_{\rm beam}$ is the number of beams covering the source extension.

\subsection{Optical}
\label{sec:opt}

\begin{figure}
\centering
\includegraphics[width=.5\textwidth]{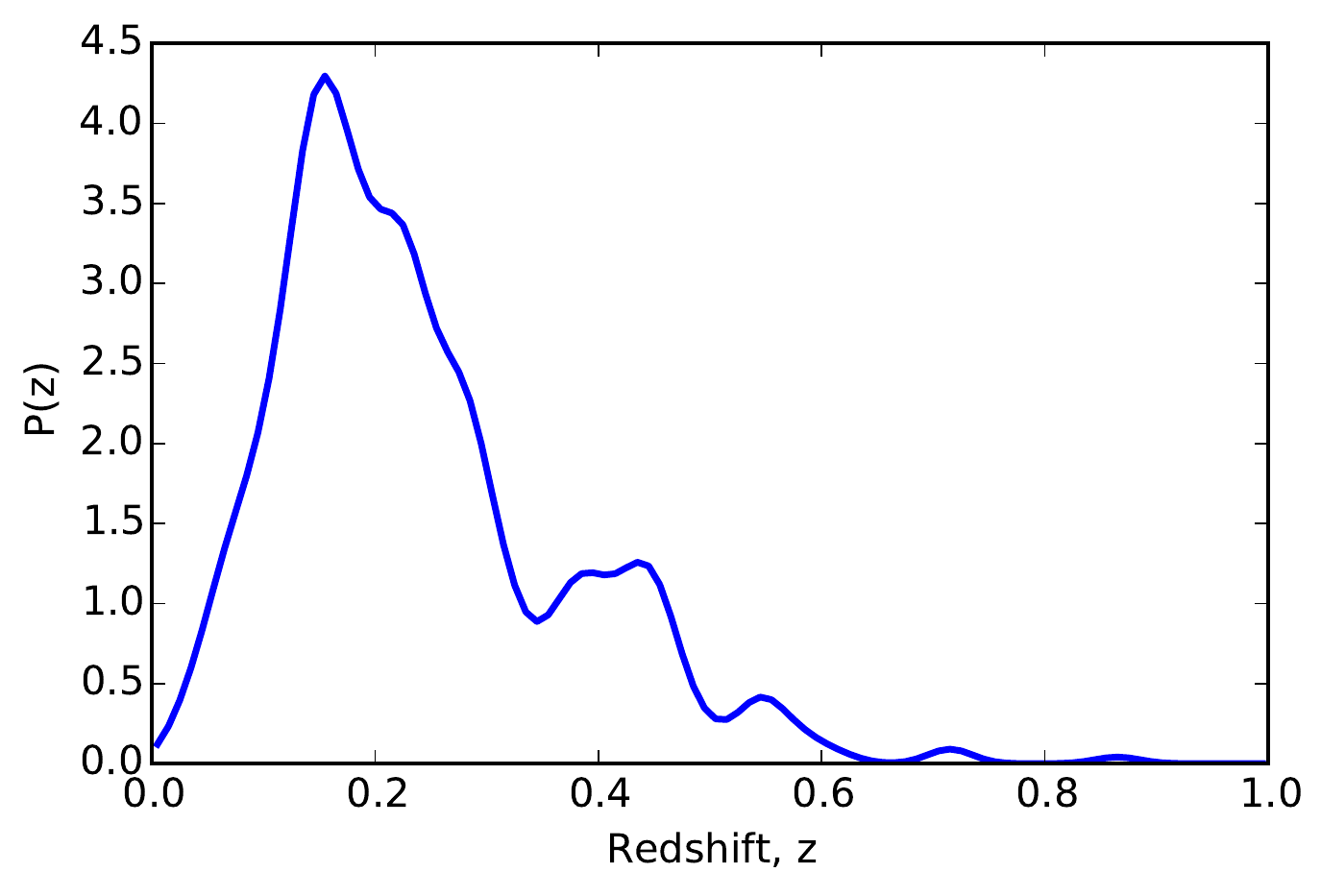}
\caption{Normalised redshift probability density function of sources with W1 and W2 magnitudes within 2-sigma of RG1 for a set of semi-analytic models from SAGE \citep{Croton2016}. The peak of the distribution is at $z=0.16$.} \label{fig:z}
\end{figure}

Although pointed optical observations of the nearby clusters PSZ1 G108.18-11.53 are available, none of them cover the region of the hosting galaxy of RG1 or RG2. The redshift of both galaxies is currently unknown, but a photometric redshift should be included in the forthcoming release of PanSTARSS. To have a rough estimation of the redshift of RG1 we compared the WISE (W1 and W2 band) fluxes of RG1 with the redshift distribution of sources in the SAGE \citep[Semi-Analytic Galaxy Evolution;][]{Croton2016} models with comparable WISE fluxes and colours (see Fig.~\ref{fig:z}). The predicted redshift probability density function, $P(z)$, was generated by averaging the redshift distributions of photometrically similar sources in 5 model lightcones with different sightlines. The model photometry was generated using the Flexible Stellar Population Synthesis stellar population models \citep{Conroy2009} and assuming a Chabrier initial mass function \citep{Chabrier2003}. Additionally, dust attenuation is incorporated into the model photometry following the method outlined in \cite{Tonini2012}. However, given the long wavelength of the filters being used, the effect of dust on the apparent magnitudes of model sources is likely very small. The most likely redshift is around $z=0.16$, but this estimation must be taken just as a qualitative indication and it will not be used to derive any important conclusion in the rest of the paper.

\subsection{X-ray}
\label{sec:xray}

\begin{figure}
\centering
\includegraphics[width=.5\textwidth]{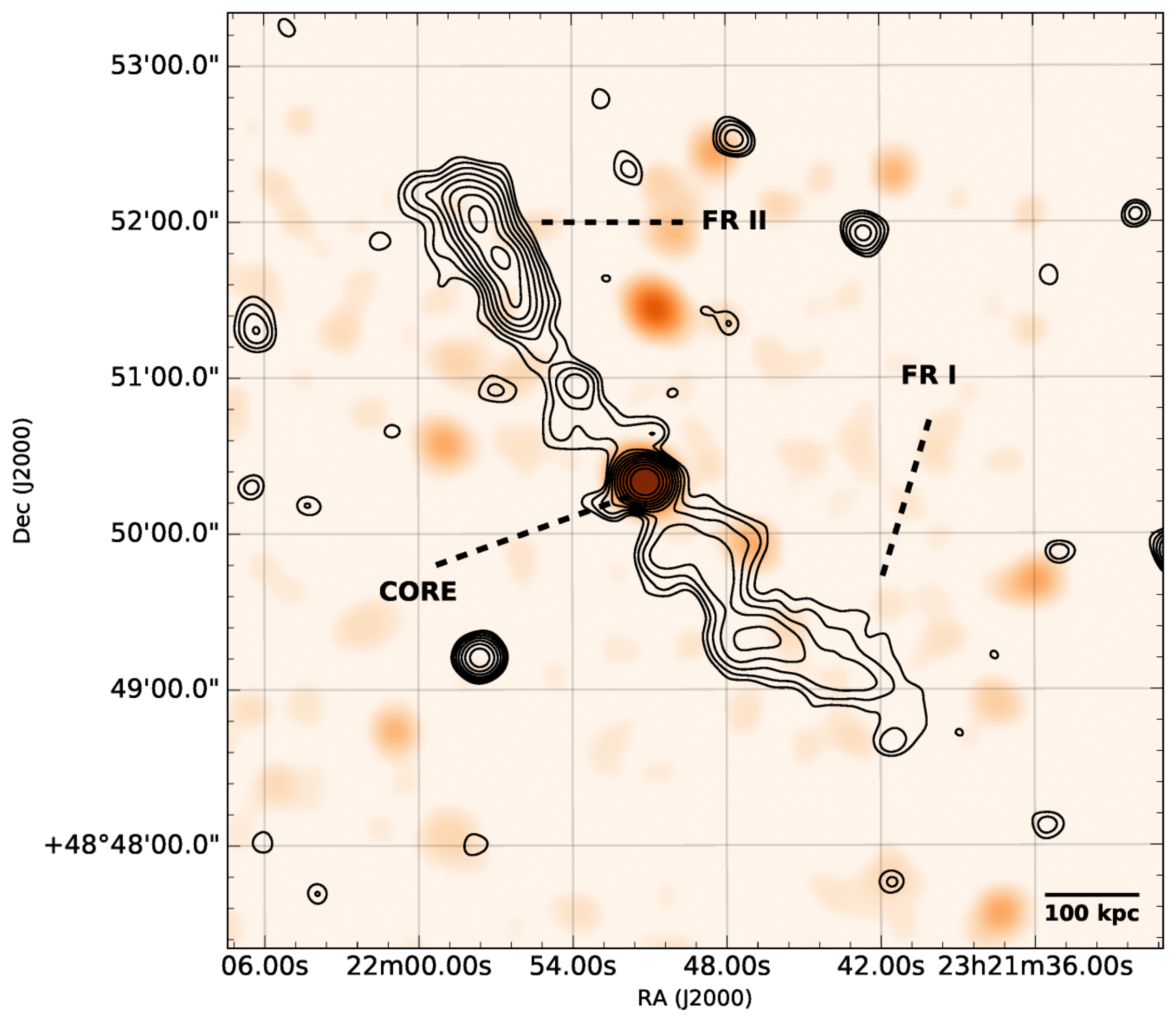}
\caption{Chandra X-ray image of the field surrounding \targetA{} (RG1). A few point sources are visible, one is co-located with the HyMoRS core, no X-ray excess is present in the location of the expected FR\,II hotspot. Radio contours are from VLA image at 1519 MHz: $(3..100) \times \sigma$, logarithmically-spaced, with $\sigma=15$~\mujybeam (beam: \beam{11}{10}).} \label{fig:VLA1500b}
\end{figure}

\begin{figure}
\centering
\includegraphics[width=.5\textwidth]{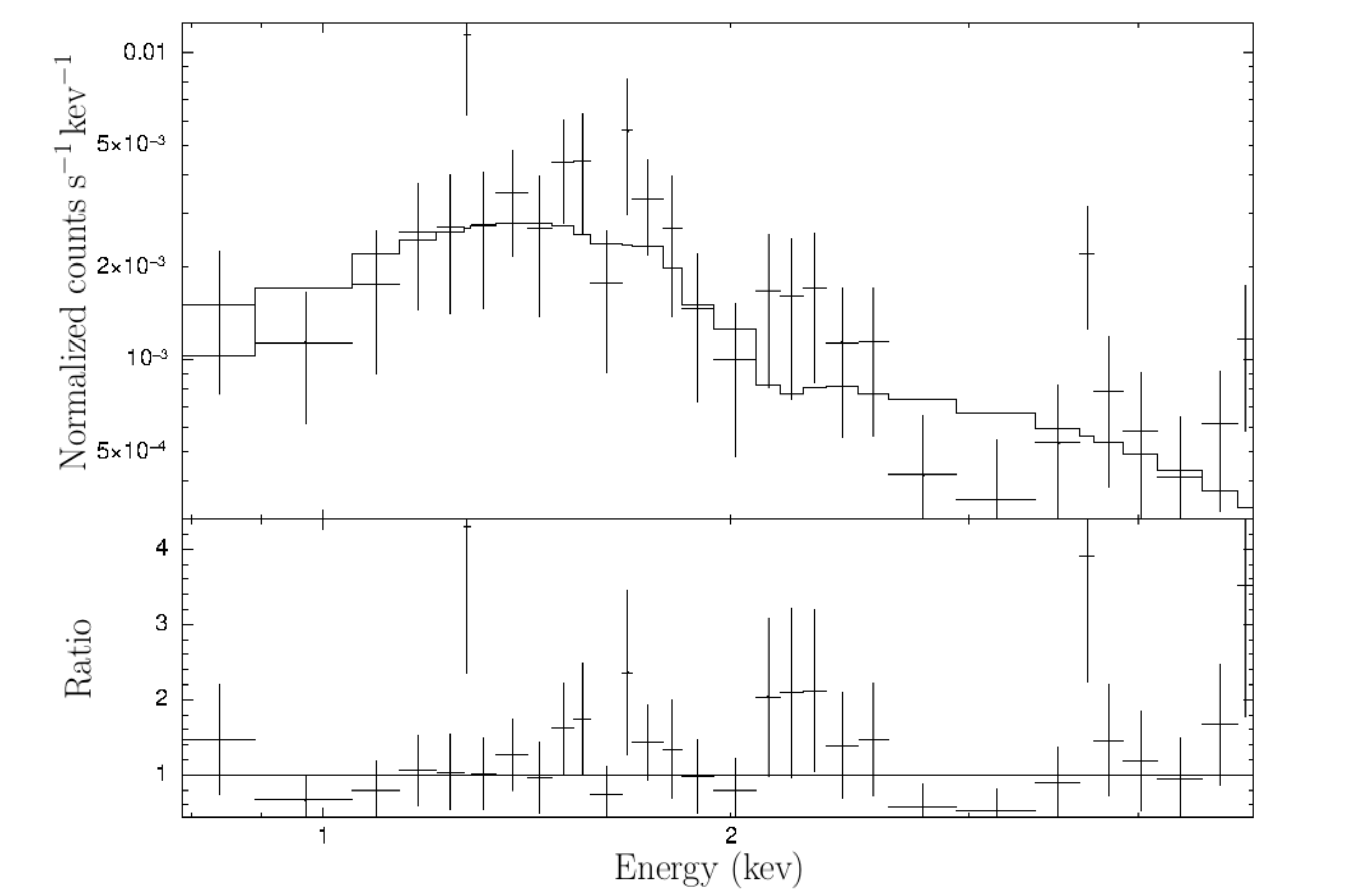}
\caption{X-ray spectra of the point source co-located with the core of RG1.} \label{fig:xraySED}
\end{figure}

A Chandra ACIS-I observation of 27~ks (obs. 801493, performed on 09/23/2014), pointed on the nearby cluster PSZ1 G108.18-11.53, has been reduced following the strategy described in \cite{vanWeeren2016}. The only significant emission that is associated with the candidate HyMoRS RG1 is co-located with the radio galaxy core (see Fig.~\ref{fig:VLA1500b}). Given the number of photons detected we were able to fit only a simple power-law. Galactic $N_H$ absorption was fixed at $1.25 \times 10^{21}\ \rm{cm}^{-2}$ \citep{Kalberla2005}. We obtained a slope of the electron distribution equals to $1.67 \pm 0.21$. The fit is shown in Fig.~\ref{fig:xraySED} and, although the noise is large, it does not show evidence for strong internal absorption.

\section{A true HyMoRS: \targetA{} (RG1)}
\label{sec:rg1}

\begin{table*}
\centering
\begin{threeparttable}
\begin{tabular}{lccccccc}
     & RA       & DEC      & Flux 323 MHz\tnote{a} & Flux 607 MHz\tnote{a} & Flux 1380 MHz\tnote{a} & Flux 1519 MHz\tnote{a} & Integrated \\
     & hh:mm:ss & dd:mm:ss & (mJy)                 & (mJy)                 & (mJy)     & (mJy)    & spectral index\tnote{b} \\
\hline
RG1 (HyMoRS) & 23:21:51.1 & 48:50:20 & $78\pm1$   & $41\pm1$     & $15.0\pm0.2$  & $15.1\pm0.3$ & $-1.10\pm0.01$ \\
RG2          & 23:21:48.1 & 48:30:04 & $41\pm1$   & $27\pm1$     & $12.1\pm0.2$  & $14.9\pm0.3$ & $-0.75\pm0.02$ \\
\end{tabular}
\begin{tablenotes}
    \item[a] Flux density calculated within the $3\sigma$ contour of the 323~MHz map.
    \item[b] Using all available frequencies.
\end{tablenotes}
\end{threeparttable}
\caption{Radio galaxies flux densities}\label{tab:RG}
\end{table*}

% IR: wise (eq.1 http://wise2.ipac.caltech.edu/docs/release/allsky/expsup/sec4_4h.html):
% 31.674 * 10**(-12.414mag/2.5) = 0.3428mJy -> nuLnu=log10(2e13*2.4e29)=42.7 with c/15micron=2e13Hz
% radio: nuLnu@178MHz -> log10(178e6*5.5e31 * (178/323)**(-1.1))=40.3
% xray: log10(1.5e-13 erg/cm/cm/s * 4 * pi * (765e6 pc)**2 in erg/s) = 43.0
RG1 is likely not part of the cluster PSZ1 G108.18-11.53 but rather a foreground object. This conclusion is based on its distance from the cluster centre and angular size. The rough redshift estimation made in Sec.~\ref{sec:opt} ($z=0.16$), also places the source in the foreground with respect to the cluster ($z=0.335$). Assuming that redshift, the radio galaxy has a luminosity at 1.4 GHz of $(1.16 \pm 0.02) \times 10^{24}$~W/Hz. According to the historical FR\,I/FR\,II luminosity division \citep{Owen1994} the source would fall in the FR\,I regime, however it is now known that both the morphological modes populate these luminosity ranges \citep{Best2009}. The luminosity in infrared (WISE at 12~$\mu$m, $\log(\nu L_\nu / {\rm erg\,s^{-1}}) = 42.7$), X-ray (Chandra $2-10$ kev, $\log(L_{\rm unabs} / {\rm erg\,s^{-1}}) = 43.0$), and radio (178 MHz, $\log(\nu L_\nu / {\rm erg\,s^{-1}}) = 40.3$) points towards a classification as low-excitation radio galaxy \citep[LERG,][]{Mingo2014}.

\begin{figure*}
\centering
\subfloat[RG1 (DSS)]{\includegraphics[width=.5\textwidth]{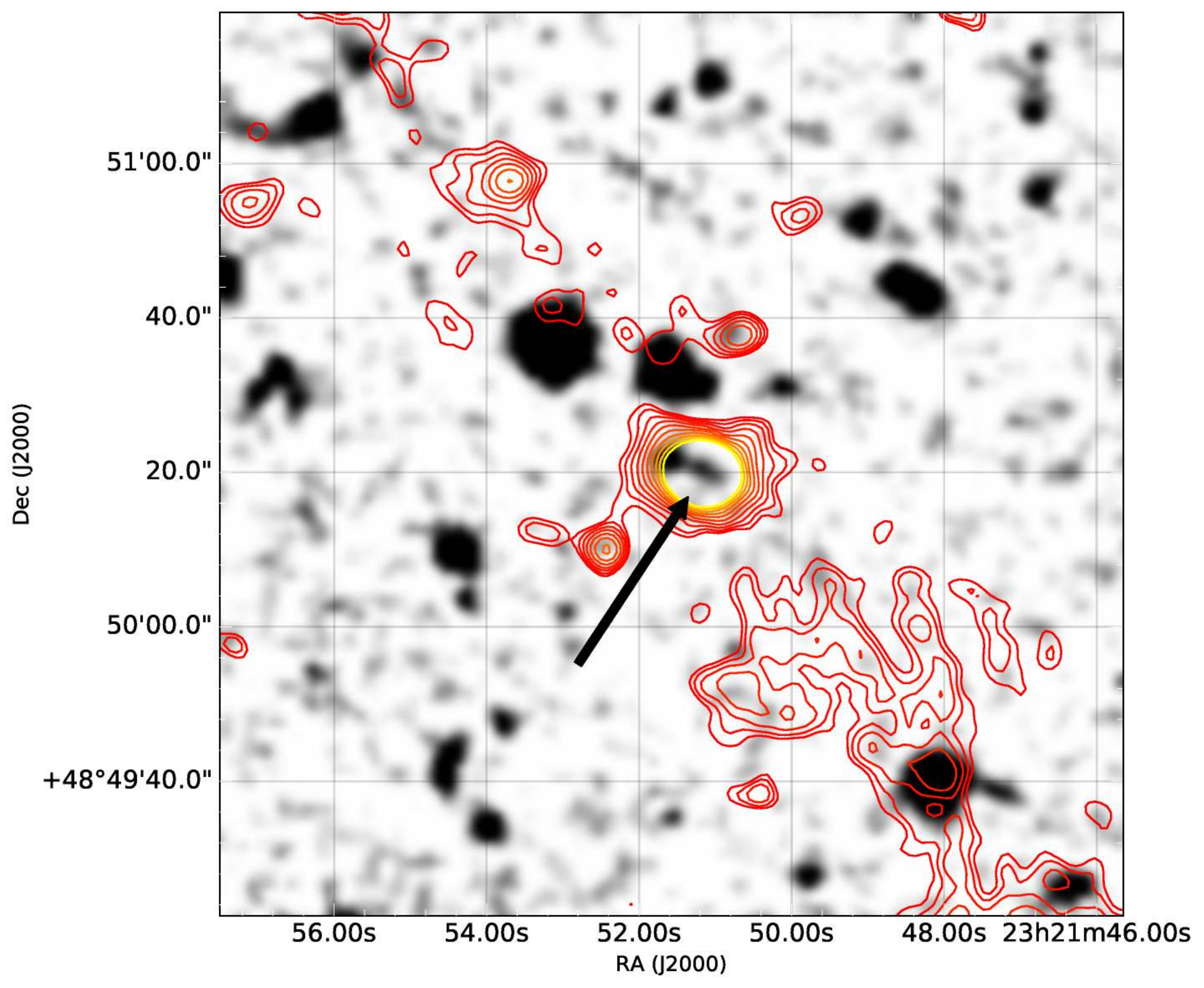}\label{fig:RG1_opt}}
\subfloat[RG2 (DSS)]{\includegraphics[width=.5\textwidth]{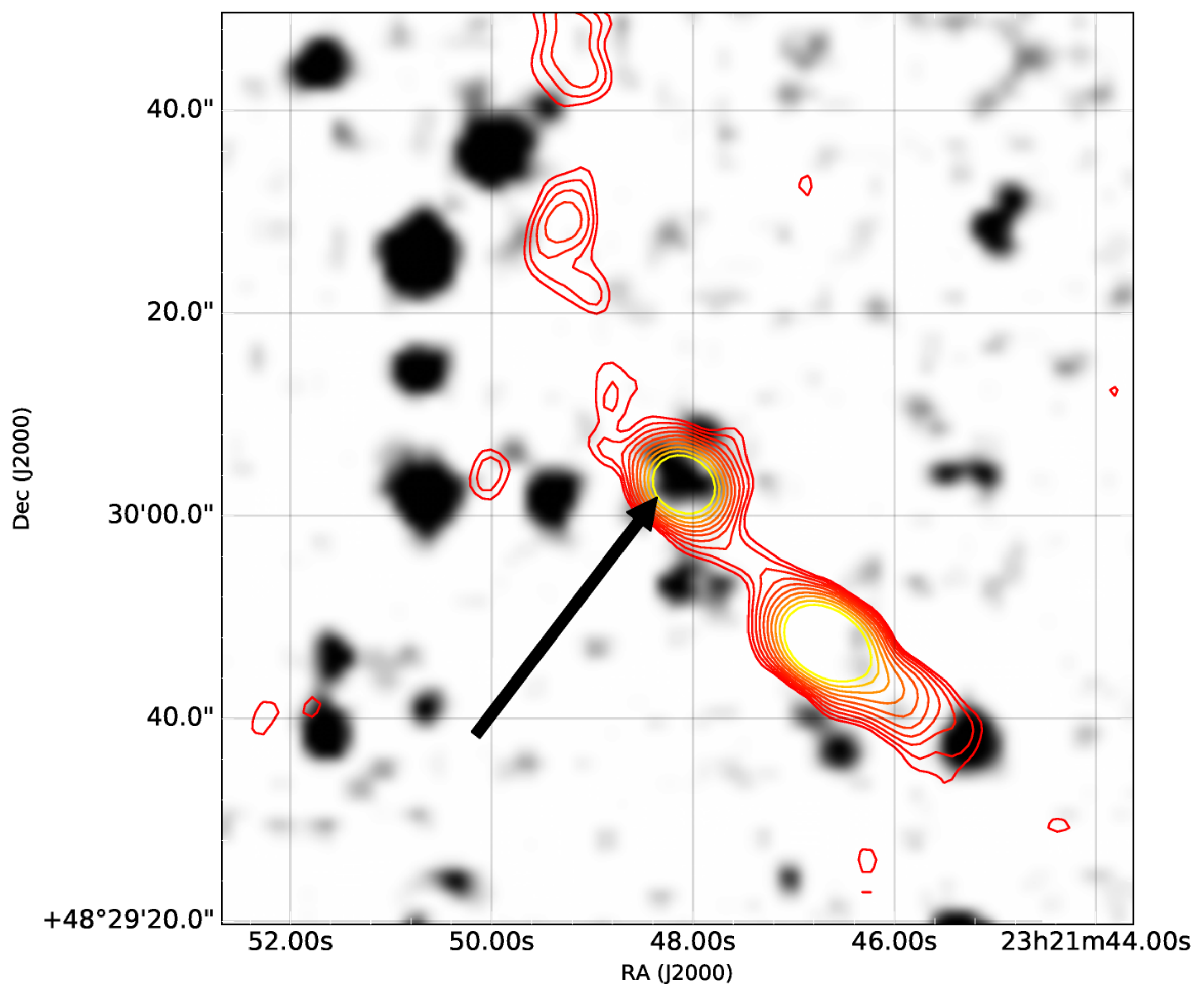}\label{fig:RG2_opt}}
\caption{DSS images (IR filter) of the cores of the HYMORS RG1 (left) and the non-HYMORS RG2 (right). Contours in the two panels are from the VLA image in Fig.~\ref{fig:radio} and  Fig.~\ref{fig:radio2} respectively. The optical counterparts are indicated by the arrow. The hosting galaxy might be in a merging state in both cases.} \label{fig:opt}
\end{figure*}

The extension of RG1 is approximately 5\arcmin, which corresponds to a projected linear size of 840 kpc at $z=0.16$. The nucleus of the source has a faint infrared WISE counterpart, an optical DSS counterpart (Fig.~\ref{fig:opt}) and an X-ray Chandra counterpart (Fig.~\ref{fig:VLA1500b}) at RA: \hms{23}{21}{51.1}, Dec: \dms{48}{50}{20}. The active galactic nucleus (AGN) is also identifiable in the spectral index map as a region with an inverted spectrum ($\alpha = +0.14 \pm 0.03$, Fig.~\ref{fig:spidx}).

The southern lobe has a clear FR\,I morphology. It is bent and does not have a clear termination point, instead it fades into noise. The surface brightness is rather uniform along the lobe (see Fig.~\ref{fig:spidxflux}). The emission shows signs of polarisation to an average level of 20\%. However, the magnetic field shows a quite disordered pattern, similar to the total intensity, which indicates a turbulent medium.

On the contrary, the northern lobe has typical characteristics of FR\,II sources. The lobe brightness is enhanced towards its termination and a discontinuity in the flux density, which increases by a factor of 3, is detected in the same region where the spectral index flattens (Fig.~\ref{fig:spidxflux}). In the location of enhanced brightness, the lobe emits 16\% polarised radiation. The magnetic field orientation is parallel to the jet direction, as seen in FR\,II sources \citep{Bridle1984}. However, no prominent single hotspot is detected in that region. We report instead three surface brightness peaks aligned with the jet which might indicate the presence of multiple hotspots \citep[as seen in other radio sources like 3C\,20, 3C\,405 and 3C\,351;][]{Hardcastle2008}. No X-ray emission is detected in the location of the hotspot, which is not uncommon and has been reported for a number of bona fide FR\,II radio galaxies \citep[see e.g.][]{Hardcastle2004}. RM-synthesis can be a useful tool to study inhomogeneous medium around HyMoRS cores. However, in this case it does not show relevant features and the measured Faraday depth in both lobes is compatible with being generated by our Galaxy \cite[$-14$~\radmm,][]{Oppermann2012}.

\subsection{Spectral analysis}
\label{sec:spidx}

Integrated flux density values and spectral index are listed in Table~\ref{tab:RG}. More interesting is to study the spectral index profiles across the source extensions as they are distinct markers of FR\,I or FR\,II sources. In radio galaxies the spectral energy distribution of the synchrotron emission is rather flat close to the location where electrons are accelerated. Synchrotron and inverse Compton losses deplete the more energetic electrons faster then the less energetic ones, steepening the spectrum with time. In the case of FR\,I sources the jet is disrupted close to the central engine and a steepening of the spectrum should be visible moving away from the core. On the contrary, in FR\,II sources the jet propagates relatively unperturbed until the ``hotspots'' where it forms shocks that re-accelerate the electrons. These electrons then flow back towards the core. As a consequence, the radio spectra is expected to be flatter close to the hotspot and steeper close to the core, where the aged plasma accumulates.

\begin{figure}
\centering
\includegraphics[width=.5\textwidth]{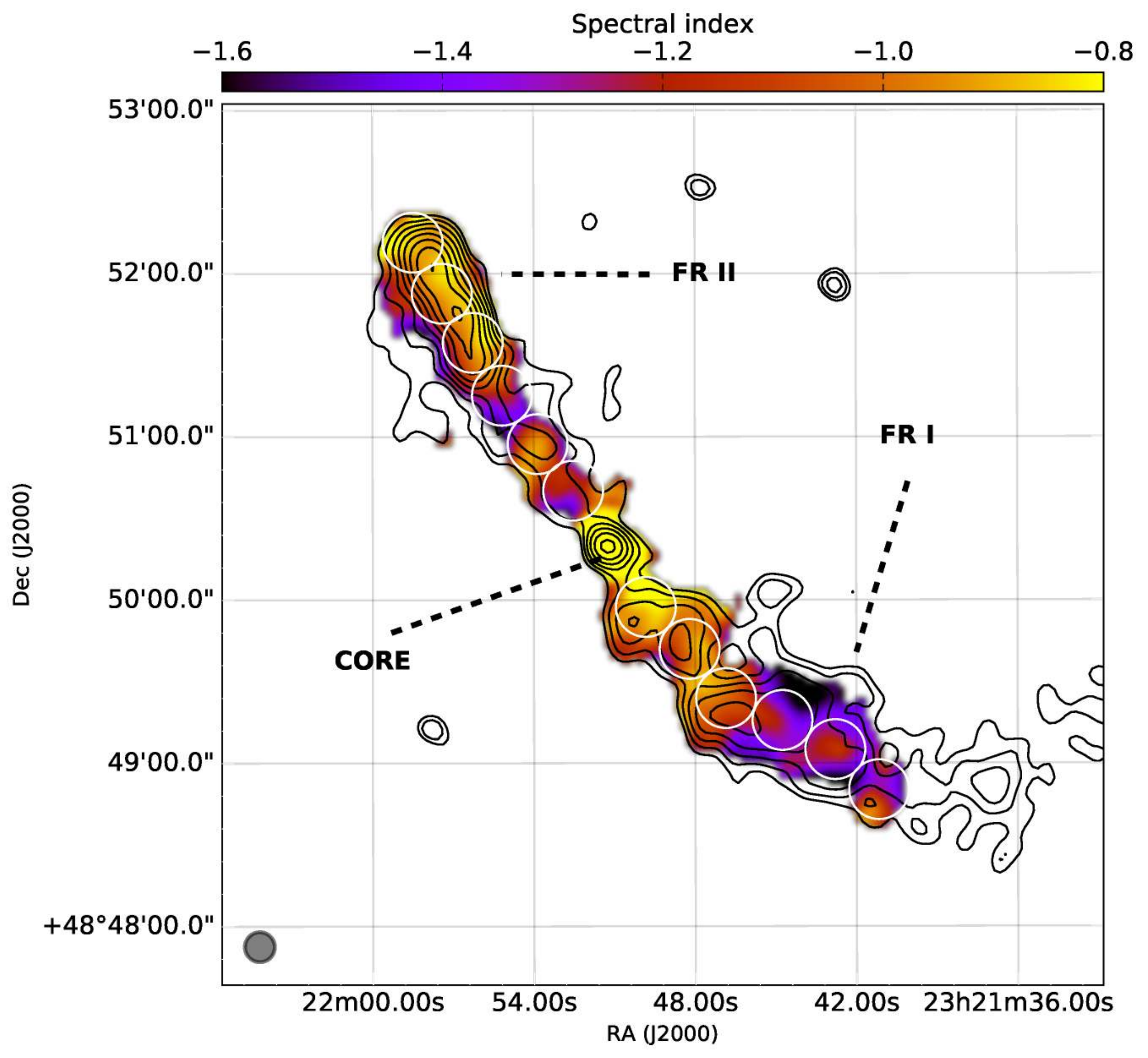}
\caption{Spectral index maps obtained from the 1519 and 323 MHz maps. Over-plotted the contours of the 323 MHz map at $3...50\times\sigma$ logarithmically spaced, with $\sigma=170$~\mujybeam and beam = \beam{18}{18} for both images. Dashed lines represent the regions used to extract the flux densities and spectral index values plotted in Fig.~\ref{fig:spidxflux}. The spectral index of the cores is rather flat, while gradients are seen in the lobes of RG1. The gradients are in-line with the definition of FR\,I and FR\,II associated with the southern and the northern lobe, respectively. On the contrary, RG2 does not show clear gradient, but the spectral index allows to identify the core position.}\label{fig:spidx}
\end{figure}

\begin{figure}
\centering
\includegraphics[width=.5\textwidth]{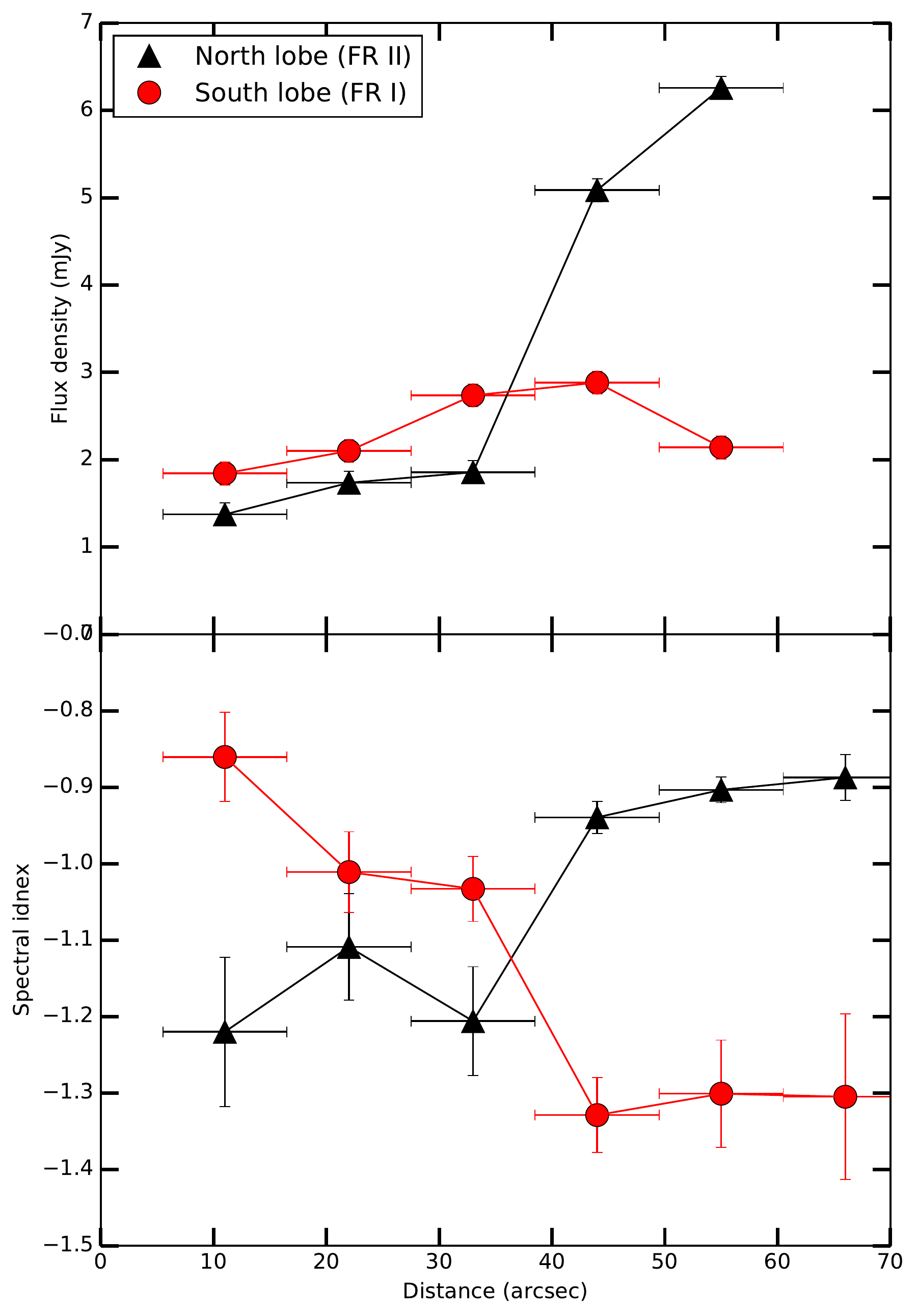}
\caption{Top: integrated flux density at 323 MHz of the regions shown in Fig.~\ref{fig:spidx}. On the $x$ axis is the distance of the region from the source core; error-bars are used to show the region extension. Bottom: integrated spectra index values of the regions shown in Fig.~\ref{fig:spidx}. The two lobes of RG1 show opposite trends as expected for HyMoRS. This is not the case for RG2, where both lobes are compatible with a constant spectral index along their extensions.} \label{fig:spidxflux}
\end{figure}

To test these scenarios, we used the observations at 323 and 1519 MHz, where the most detailed radio images were obtained. The datasets were re-imaged tapering the data so to obtain the same resolution ($\sim 11\arcsec$) in both maps. GMRT at 325 MHz and VLA at L-band in C-configuration cover the same minimum $uv$-distance in wavelength, this means that the two interferometers are sensitive to the same maximum angular scale. This minimises the bias due to the different $uv$-coverages that is unavoidable when doing aperture synthesis of extended sources. Of the two images, we retained only pixels that were above $3\sigma$ in each map. We divided the source lobes into circular regions with a diameter twice the synthesised beam FWHM (see Fig.~\ref{fig:spidx}). For each region we computed the integrated spectral index using a bootstrap least squares method to include flux density errors. The results of this analysis are shown in Fig.~\ref{fig:spidxflux}. RG1 shows opposite trends in the two lobes. The northern lobe has a steep spectral index ($\alpha = -1.2 \pm 0.1$) close to the core which flattens in the outer region ($\alpha = -0.88 \pm 0.03$) due to particle reacceleration in the hot-spot and back-flow of aged electrons towards the nucleus. The southern lobe has an opposite trend: the jet is disrupted close to the core and the electrons' spectra steepens from $\alpha = -0.86 \pm 0.06$ to $\alpha = -1.3 \pm 0.1$ due to synchrotron and inverse Compton looses as the distance from the core increases.

Morphology, polarisation and spectral index characteristics all point to the classification of RG1 as a genuine HyMoRS.

\section{Classification risks: \targetB{} (RG2)}
\label{sec:rg2}

In this section we present the analysis of another candidate HyMoRS that cannot be classified as such. This case exemplifies the danger of morphology-based classification of hybrid radio galaxies.

The core of RG2 is identifiable by a flat spectral index ($\alpha = -0.01 \pm 0.06$) and by a counterpart visible in the WISE and DSS images at RA: \hms{23}{21}{48.1}, Dec: \dms{48}{30}{04} (Fig.~\ref{fig:RG2_opt}). The northern lobe of the source has an FR\,I morphology and it bends in two locations. Its spectral index is quite uniform along its extension ($\alpha \simeq -1$).

A compact emission close to the core is the brightest region of the source. It has a rather flat spectral index of $\alpha = -0.62 \pm 0.05$ and it is polarised to the level of $\sim16\%$. The polarisation angle suggests a magnetic field oriented perpendicular to the jet direction as in the case of the initial part of the jet in FR\,I sources. We interpret this emission as beamed radiation from an incoming jet.

\begin{figure*}
\centering
\includegraphics[width=.49\textwidth]{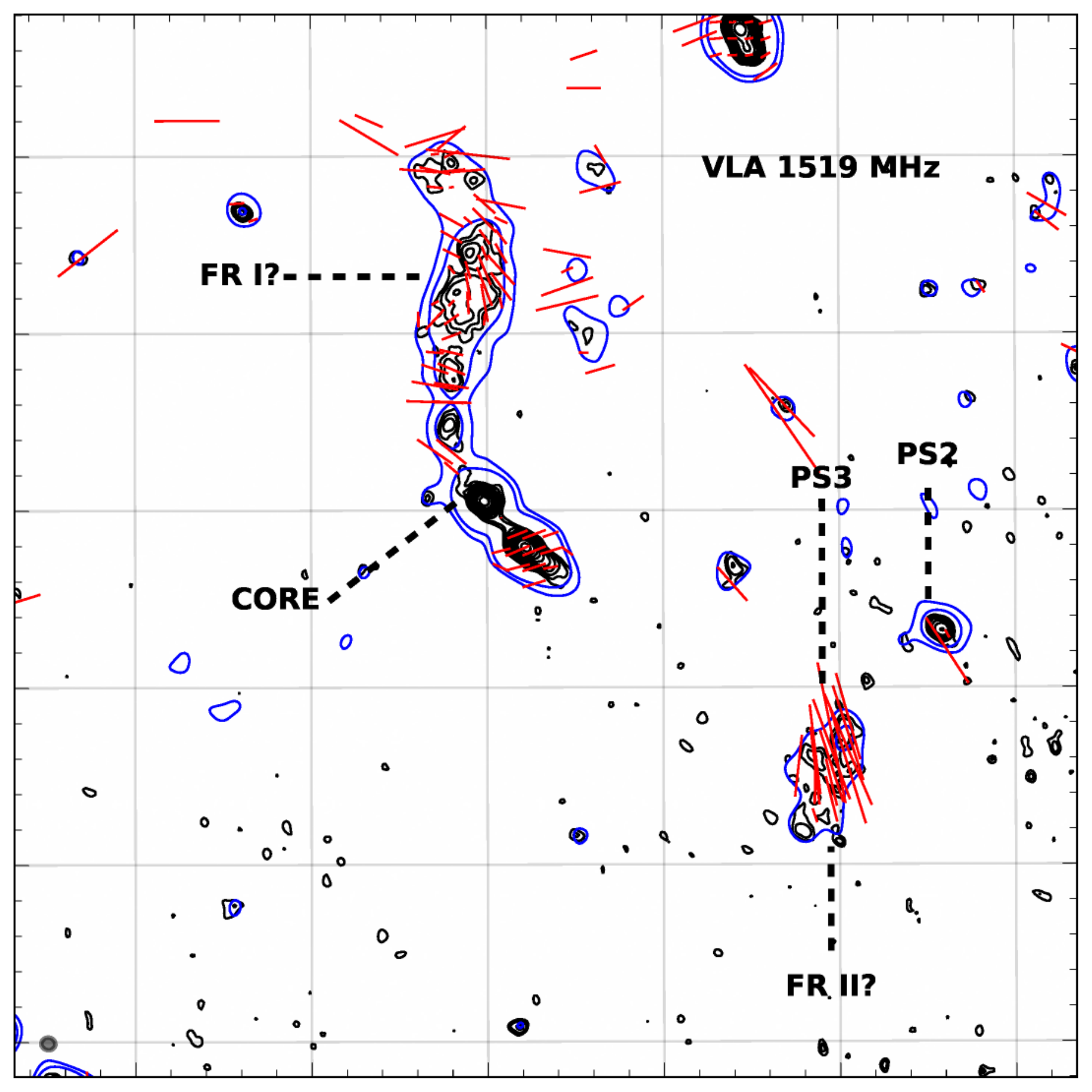}
\includegraphics[width=.49\textwidth]{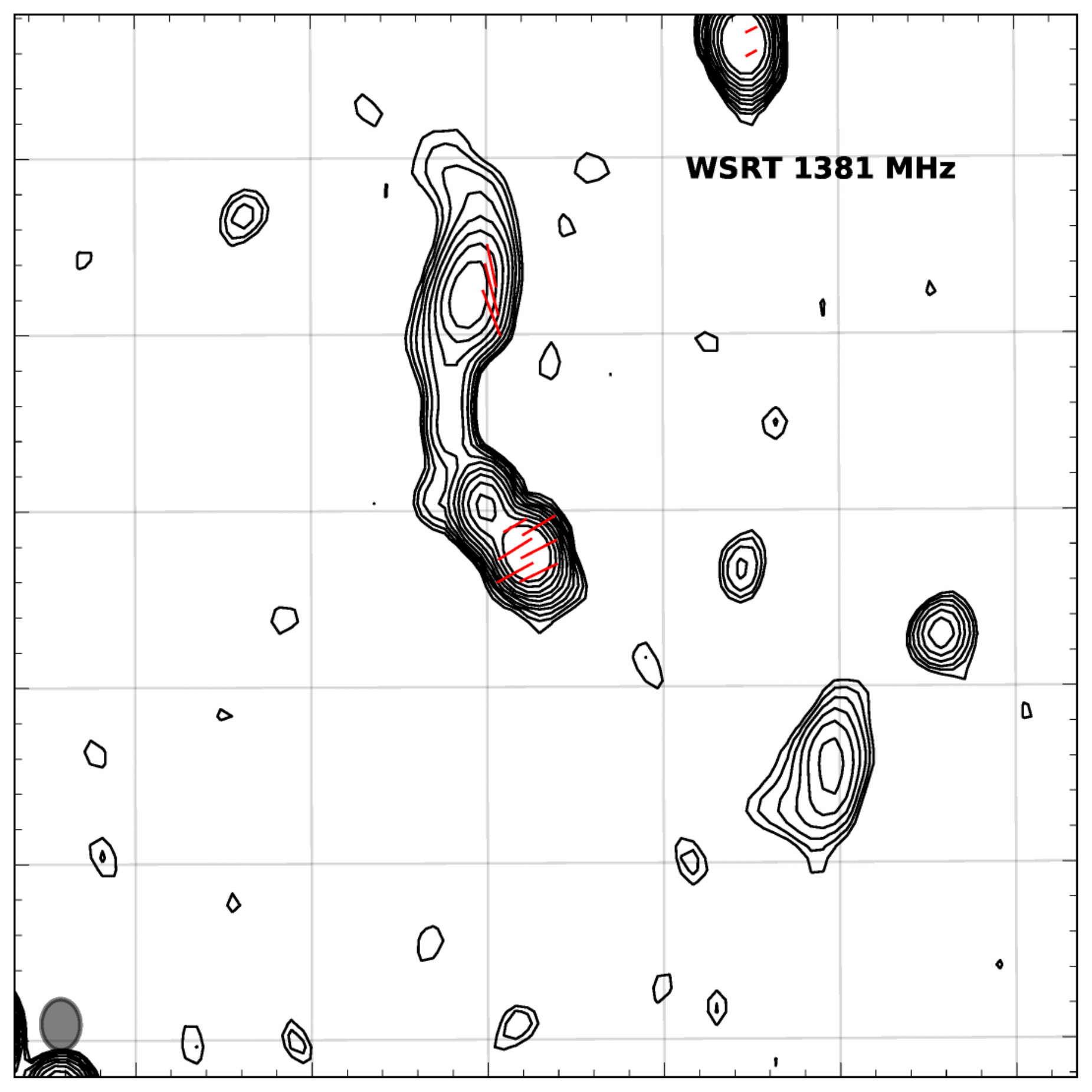}
\caption{Radio images of RG2, a beamed FR\,I radio galaxy that cannot be classified as HyMoRS. Left: black contours are from VLA data (1519~MHz) at $3...80\times\sigma$ logarithmically spaced, with $\sigma=20$~\mujybeam, beam = \beam{4.8}{4.3}. Blue contours are $(3,6)\times\sigma$ from a low-resolution images ($\sigma = 35$~\mujybeam, beam = \beam{11}{11}). Right: contours from WSRT data (1381~MHz) at $3...50\times\sigma$ logarithmically spaced, with $\sigma=37$~\mujybeam, beam = \beam{17.3}{13.3}. In both images, vectors trace the B-modes of polarised emission.} \label{fig:radio2}
\end{figure*}

Low surface brightness emission is present 2\arcmin{} south-west of the core. This emission might be part of the source lobe. There is no an apparent connection between this emission and the core-jet region, as has been seen in FR\,II lobes. However, the putative ``lobe'' lacks a compact hotspot and its spectral index is rather steep ($\alpha \lesssim -1$). We note the presence of a compact, flatter spectrum source (PS2 in Fig.~\ref{fig:radio2}, $\alpha = -0.77 \pm 0.06$) with no WISE nor DSS counterpart on the lobe's west side. A point source with a WISE counterpart is also present on the northern edge of this extended emission (PS3 in Fig.~\ref{fig:radio2}). A alternative interpretation could be that this extended emission is not part of the RG2 source but it is a lobe related with the compact source PS3.

Although the source is difficult to interpret and probably strongly beamed, the data available to us do not suggest RG2 to be a hybrid radio galaxy. However, a low resolution image (see Fig.~\ref{fig:radio2}) without spectral index or polarisation information (as a radio survey image) could have led to the HyMoRS classification of RG2 based on its morphology.

\section{Discussion and Conclusions}
\label{sec:conclusions}

Some recent works support the idea that the FR\,I/FR\,II dichotomy might be related to environmental effects. In fact, in their central region, HyMoRS appear as FR\,II sources. This finding is supported by VLBI observations of the central 10 kpc \citep{Ceglowski2013} and by measuring X-ray -- radio scaling relations of the nuclear emission \citep{Miller2009}. In this scenario, one side of the source may decelerate and turn into an FR\,I type. A theoretical framework that can explain this behaviour was developed by \cite{Meliani2008}, where the jet deceleration is caused by a density jump in the external medium. According to \cite{Perucho2012}, the FR\,II type jet can also be disrupted by the growth of helical instability originated by the interaction with the surrounding media. It is interesting to notice that the magnetic field on the FR\,I side of RG1 seems indeed to trace an helical flow which might have disrupted the original FR\,II straight jet. Alternatively, the jets of FR\,II sources might be eroded by the entrainment due to the interactions with their surrounding lobes \citep{Wang2011}.

Other theories link the arising of an FR\,I or an FR\,II to the nucleus properties \citep{Wold2007}. However, X-ray and optical spectroscopic studies \citep{Buttiglione2010,Best2012} support the division of radio sources based on line ratios (low excitation galaxies or LEG and high excitation galaxies or HEG) rather then their observed morphology, and link the discussion to the properties of the central engine: black hole spin or mode of accretion. In this classification, weak FR\,IIs (LEGs) fueled via radiatively inefficient flows at low accretion rates are grouped together with FR\,Is \citep[e.g.][]{Best2012}.

The picture that seems to arise is that all radio sources begin life as FR\,II sources, with collimated jets, but that some sources do not survive as such to large sizes, getting disrupted into FR\,Is \citep{Kaiser2007,Kunert-Bajraszewska2010}. Therefore, the FR\,I/FR\,II nature of a radio galaxy must be related with both the central engine properties and accretion mode, which drives the jet power, as well as with the surrounding environment, which opposes the jet advancement. In this picture, HyMoRS are FR\,II sources with a jet power not much higher than that of an FR\,I. As a consequence, a variation in the environment around one of the two jets could cause its disruption and transformation in a turbulent FR\,I lobe.

In this work we showed that HyMoRS have a complete bimodal nature under the point of view of morphology, polarisation and spectral properties, showing no apparent difference between a ``standard'' FR\,I (FR\,II) lobe and an HyMoRS FR\,I (FR\,II) lobe. We presented multi-frequency, polarimetric observation of two candidate HyMoRS. \targetA{} (RG1) appears to be a rare example of an intrinsically-true hybrid source as indicated by morphology, spectral index and polarisation analysis. On the other hand \targetB{} (RG2) appears strongly beamed and does not show any hotspot on the putative FR\,II side and cannot be classified as an HyMoRS. Our work solidifies the importance to analyse multi-frequency, high-resolution and polarisation maps to accurately distinguish HyMoRS. More insight on the HyMoRS phenomenon will come from the analysis of larger samples that will be assembled soon thanks to high resolution, high survey-speed telescopes such as LOFAR and SKA.

\section*{Acknowledgements}
The author would like to thank Marcus Br\"uggen, Philip Best, Reinout van Weeren and Kenneth Duncan for the discussions and suggestions.
%We thank the anonymous referee for her/his useful comments.

% GMRT
We would like to thank the staff of the GMRT that made these observations possible. GMRT is run by the National Centre for Radio Astrophysics of the Tata Institute of Fundamental Research.

% DS9
%This research has made use of SAOImage DS9, developed by Smithsonian Astrophysical Observatory.

% NRAO
The National Radio Astronomy Observatory is a facility of the National Science Foundation operated under cooperative agreement by Associated Universities, Inc.

% SDSS
%Funding for SDSS-III has been provided by the Alfred P. Sloan Foundation, the Participating Institutions, the National Science Foundation, and the U.S. Department of Energy Office of Science. The SDSS-III web site is \url{http://www.sdss3.org}.

% NED
%This research has made use of the NASA/IPAC Extragalactic Database (NED) which is operated by the Jet Propulsion Laboratory, California Institute of Technology, under contract with the National Aeronautics and Space Administration.

% ADS
%This research has made use of NASA's Astrophysics Data System.

\bibliographystyle{mn2e}
\bibliography{HYMORS}
\bsp

\label{lastpage}

\end{document}